%% file: redux_revised4.tex
\documentclass[useAMS,usenatbib]{mn2e}
\usepackage{times}
\usepackage{amsmath}
\usepackage{graphicx} 
\usepackage{multicol}
\input{chocros}

\title[Sensitivity and Variability Redux]{Sensitivity and Variability
  Redux in Hot-Jupiter Flow Simulations}

\author [J.~Y-K.~Cho, I.~Polichtchouk \&
H.~Th.~Thrastarson]{J.~Y-K.~Cho,$^{1,2}$\thanks{On leave from QMUL;\ \
    \ Email: J.Cho@qmul.ac.uk} I.~Polichtchouk,$^{1,3}$
  H.~Th.~Thrastarson,$^4$\\
  $^1$ School of Physics and Astronomy, Queen Mary
  University of London, London E1 4NS, UK\\
  $^2$ Institute for Theory and Computation, Harvard University,
  Cambridge, MA 02138, USA\\
  $^3$ Department of Meteorology, University of Reading, Reading RG6 6BB, UK\\
  $^4$ Jet Propulsion Laboratory, California Institute of Technology,
  Pasadena, CA 91109, USA}

\begin{document}

\date{Accepted yyyy mmm dd. Received yyyy mmm dd; in original form
  yyyy mmm dd}

\pagerange{\pageref{firstpage}--\pageref{lastpage}} \pubyear{yyyy}

 \maketitle

\label{firstpage}

\begin{abstract}
  We revisit the issue of sensitivity to initial flow and intrinsic
  variability in hot-Jupiter atmospheric flow simulations, originally
  investigated by \citet{Choetal08} and \citet{ThraCho10}.  The flow
  in the lower region ($\sim$1 to $20$~MPa) `dragged' to immobility
  and uniform temperature on a very short timescale, as in
  \citet{LiuShow13}, leads to effectively a complete cessation of
  variability as well as sensitivity in three-dimensional (3D)
  simulations with traditional primitive equations.  Such momentum
  (Rayleigh) and thermal (Newtonian) drags are, however, ad hoc for 3D
  giant planet simulations.  For 3D hot-Jupiter simulations, which
  typically already employ strong Newtonian drag in the upper region,
  sensitivity is not quenched if only the Newtonian drag is applied in
  the lower region, without the strong Rayleigh drag: in general, both
  sensitivity and variability persist if the two drags are not applied
  concurrently in the lower region.  However, even when the drags are
  applied concurrently, vertically-propagating planetary waves give
  rise to significant variability in the $\sim$0.05 to $0.5$~MPa
  region, if the vertical resolution of the lower region is increased
  (e.g. here with 1000 layers for the entire domain).  New
  observations on the effects of the physical setup and model
  convergence in `deep' atmosphere simulations are also presented.
\end{abstract}

\begin{keywords}
  hydrodynamics -- turbulence -- methods: numerical -- planets:
  atmospheres.
\end{keywords}

\section{Prolegomenon}\label{proleg}

Many studies have explored the effects of initial condition and
resolution or dissipation in hot-Jupiter atmospheric flow simulations
\citep[e.g.][]{Choetal08,ThraCho10,Hengetal11,ThraCho11,PoliCho12,
  Bendetal12, LiuShow13, Polietal14}.  \citet{ThraCho10} [hereafter
TC], in particular, have emphasised the nonlinear effect of initial
jet configuration on the subsequent evolution, calling attention to it
as a source of uncertainty for {\it quantitative} predictions.
`Quantitative' here means such things as the precise location of hot
regions and the three-dimensional (3D) shape and magnitude of vortices
and jets, which affect the {\it directly-coupled} temperature
distribution as well as the wave momentum and energy depositions.
Similar emphasis has also been made by \citet{Choetal08} in the
two-dimensional (2D) context with initial turbulent eddies, in which
strength of the eddies affected the final mean flow configuration.  In
this paper we revisit the effects with the commonly-used traditional
(i.e. hydrostatic) primitive equations with extended, `deep' vertical
domain.\footnote{The discussion here, however, is also germane to
  understanding results from simulations with the compressible,
  non-hydrostatic, 3D Navier--Stokes equations
  \citep[e.g.][]{DobbLin08,Maynetal14}.}

Recently, \citet{LiuShow13} [hereafter LS] have performed a set of 3D
traditional primitive equations simulations with specified thermal
(Newtonian) and momentum (Rayleigh) drags and assert that TC observe
sensitivity because they fail to meet a presumptive setup criteria for
`real' hot-Jupiters (LS, p.\ 48).\footnote{LS also emphasise that the
  equilibrium temperature and relaxation time profiles in TC are
  constant.  However, as stated in TC and explicitly shown in
  \citet{Polietal14}, the sensitivity and variability persist under
  vertically-varying equilibrium temperature distribution and
  relaxationtimescale that match closely with those used in LS, {\it
    over the domains considered in TC and Polichtchouk et al. (2014)}.}
In their work, LS also advocate the use of a strong Rayleigh drag in a
fiducial region (denoted as $\scrD$ in this paper) at the bottom of
the modelled atmosphere.  Here by `strong' we mean a specified
Newtonian or Rayleigh drag with a timescale of few (or less) planetary
rotations; and, $\scrD$ is defined as the pressure range, $p =
[1,20]$~MPa\footnote{1~MPa = 10~bars}, where both drags are `turned
on' in LS.  In this paper, we clarify the situation: the insensitivity
occurs in LS because the combined, strong Newtonian and Rayleigh drags
suppress effectively {\it all} dynamics, everywhere in the domain,
except for laminar high-speed jets.  Quantitative features of the
flow, as defined above, are sensitive to initial conditions when the
strong drags are not applied.

Rayleigh drag, which is often used to crudely represent a physical
boundary layer in 3D simulations of large-scale flows
\citep[e.g.][]{HeldSuar94}, is implausible for the $\scrD$ region of a
giant planet without a `solid' surface or with one located deep in its
interior -- particularly if the drag is a strong
one.\footnote{Recently, \citet{SchnLiu09} have put forth a weak
  Rayleigh drag representation of `magnetohydrodynamics-induced drag'
  effect, but the representation is at present not widely accepted.}
The use of such a drag in $\scrD$ is notable given that past
hot-Jupiter studies without the drag have already reported a lack of
sensitivity and variability \citep[e.g.][]{Showetal09, CoopShow05}:
hence, it is not required per se to demonstrate insensitivity.
Nonetheless, to elucidate the effects of the new setup with strong
drags, we set up here simulations as in LS and cross-check the results
with several different codes.  In doing so, we explicitly demonstrate
that the sensitivity is quenched only under the application of an
unphysical, strong Rayleigh drag and that such a drag is not necessary
for attaining equilibration if a strong Newtonian drag is already
applied.\footnote{The latter point was not made explicit in TC and
  some confusion appears to exist concerning `equilibration' under
  strong, planetary-scale Newtonian drag -- perhaps unduly influenced
  by barotropic non-divergent (i.e. incompressible 2D) turbulence
  formalism, which does not involve the thermodynamic fields.}  In any
case, `equilibration' reached -- with or without the drag -- is best
regarded as heuristic and should not be taken too literally in the
present setup because unphysical supersonic flow results from it.
Even with both drags applied, variability and weak sensitivity persist
if vertical resolution of the $\scrD$ region is increased in the
simulations.

\section{Setup}\label{setup}

TC and LS solve the traditional primitive equations
\citep[e.g.][]{Holt04} with the same, or effectively the same,
boundary conditions (zero `vertical velocity' -- i.e. `free slip' --
at the top and bottom of the domain).\footnote{MITgcm has an option to
  allow the bottom pressure surface to be free (i.e. nonzero vertical
  velocity).  The state of this option is not specified in LS; but,
  the results are not noticeably affected by the option for the setup
  as in LS.  However, in general, energy is added and stability is
  affected with the option on.  The option is off in all the
  simulations discussed in this paper.}  However, the basic numerical
algorithm, treatment of explicit viscosity and vertical coordinate are
all very different: in TC pseudospectral and superviscosity with CAM
\citep{Coll04} in $\eta$-coordinate and in LS finite-volume and
moderate-order Shapiro filter with MITgcm \citep{Adcr12} in
$p$-coordinate, with the vertical layers equally spaced in $\log(p)$.
In this study, we solve the same equations with the free-slip boundary
conditions; but, unlike in TC and LS, we employ {\it both} the
pseudospectral and finite-volume models.  While the latter is the same
model, grid configuration (cubed-sphere) and vertical coordinate used
in LS, the former is different than the model used in TC: the BOB
model \citep{Scotetal04} in $p$-coordinates, with vertical layers
equally spaced in $\log(p)$ and~$p$, is used here.  BOB is purposely
chosen to provide an independent check of TC's results, as well as
those of LS.  Note that, in general, models employing the
pseudospectral algorithm converge much faster with resolution than
models using finite-volume algorithms \citep[see e.g.][as well as the
discussion below]{Polietal14}.

In all three studies (TC, LS and the present one), planetary parameter
values characteristic of the hot-Jupiter HD209458b are used (see TC
for the values).  In this paper, all times and lengths are scaled by
the planetary rotation period ($\tau\! =\! 3.024\! \times\! 10^5$~s)
and radius (${\cal L}\! =\! 10^8$~m), respectively -- unless units are
explicitly given.  Hence, $t\! =\! 1$ corresponds to 1~HD209458b day.
Also in all three studies, the thermal forcing is crudely represented
by a relaxation (drag) to a specified `equilibrium' temperature
distribution $T_e(\lambda,\phi,p)$ with a drag-time profile
$\tauN(p)$; here $\lambda$ is longitude and $\phi$ is latitude.

Note that $T_e$ and $\tauN$ are not same in TC and LS, as stressed by
LS.  But, they are not very different: the difference is no more than
few percent over the model domain in some of the simulations reported
in TC and all of the forced simulations in \citet{Polietal14}, which
uses one of the vertically-varying profiles from the TC study.  In the
present study, the profile $\, T_e(p)|_{\lambda,\phi}\, $ is also not
precisely (but essentially) same as that in LS: the profile in LS is
represented as continuous, piece-wise linear functions in three
distinct $p$ regions, as $\tauN$ is in LS.\footnote{N.B. $\tauN$ here
  is `$\tau_{\rm rad}$' in LS.}  Also, our $\tauN$ in $\scrD$ is
$10^6$~s ($= 3.3$ scaled), rather than $10^7$~s, independent of $p$.
The smaller value is as in Fig.~6 of LS, and in simulations presented
in most of their discussion.  However, with both drags applied in
$\scrD$, insensitivity to initial flow does not depend on whether the
$p$-independent $\tauN$ in $\scrD$ is $10^6$~s or $10^7$~s.  In fact,
as demonstrated below, none of the above differences in the physical
setup are pertinent to the question of sensitivity: only the Rayleigh
drag, as employed in LS, is pertinent.

The crucial Rayleigh drag is characterised by the drag-time profile
$\tauR(p)$.  Note that, in all the simulations discussed in this
paper, the Rayleigh drag is applied only in the vertical region,
$\log(p)\! =\!  [5.0,7.3] = [5.0,6.0]\!\cup\!\scrD$, with $\tauR(p)$
decreasing exponentially with $\log(p)$ -- i.e.  linearly with $p$.
The specification is as in LS.  When the drag is not applied in
$\scrD$, it is not applied anywhere in the domain.  The strength of
the $p$-dependent $\tauR$ is characterised by
$\min\{\tauR(p\!\in\!\scrD)\} = \tauR(20~\mbox{MPa})$, and denoted
$\tauRB$ here.

\section{Results}\label{results}

\begin{figure*}
  \centerline{\includegraphics[scale=.94]{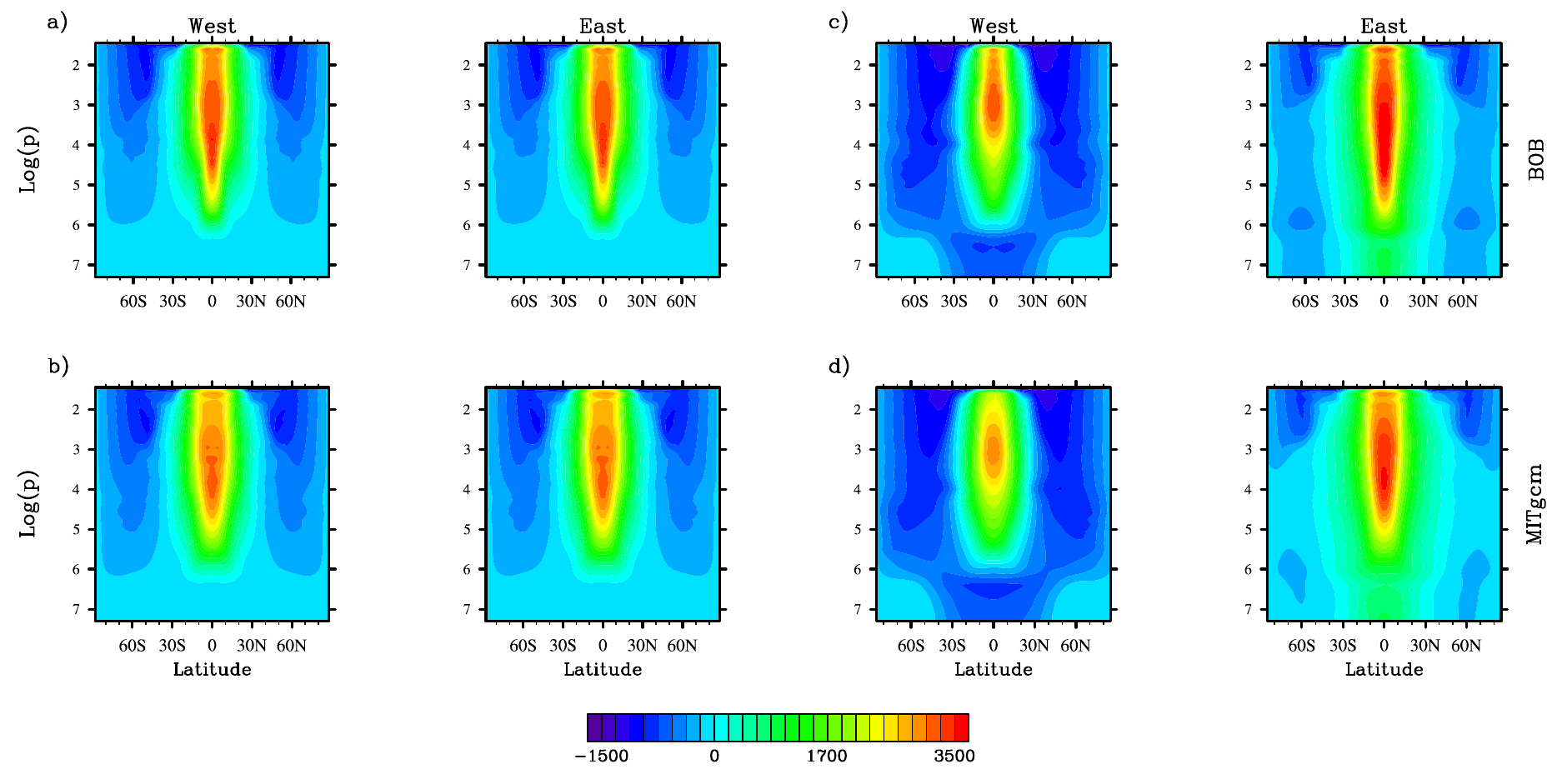}}
  \caption{Time-averaged (over $t = [250,300]$), zonal-mean zonal
    velocity $\overline{u}^\ast(\phi,p)$, where $\phi$ is latitude and
    $p$ is pressure, from simulations initialised with $\pm
    1000$~m~s$^{-1}$ (East/West) equatorial jet.  Simulations in a)
    and c) are performed with BOB and in b) and d) are performed with
    MITgcm in cubed-sphere grid.  All simulations are performed with
    vertical layers equally-spaced in $\log(p\,[{\rm Pa}])$.  In a)
    and b), $\tauN = 3.3$ independent of $\log(p)$ and $\tauR$ in
    $\scrD$ decreases exponentially with $\log(p)$ for $p > 0.1$~MPa
    with $\tauRB\!  =\!  3.3$.  In c) and d), both $\tauN,\tauR\!
    \rightarrow\!  \infty$ in $\scrD$ (i.e. no drags in $\scrD$);
    hence, $\tauRB\!  \rightarrow\!  \infty$.  When both drags are
    applied, $\overline{u}^\ast$ is essentially same, irrespective of
    the initial jet direction [a) and b)]; when both drags are not
    applied, $\overline{u}^\ast$ is strongly dependent on the jet
    direction~[c) and d)].}
  \label{fig1}
\end{figure*}

Fig.~\ref{fig1} summarises one of the basic results of this paper:
without the ad hoc strong drags applied in $\scrD$, sensitivity to
initial flow is robust.  The figure shows time-averaged, zonal-mean
zonal velocity $\overline{u}^\ast(\phi,p)$ from four pairs of
simulations with identical setup -- except for the direction of the
initial, barotropic zonal jet (see TC for details of the jet profiles)
and the presence or absence of both Newtonian and Rayleigh drags
in~$\scrD$.  In Fig.~\ref{fig1}a and \ref{fig1}b, $\tauN$ and $\tauR$
are as in the simulations of LS that apply both drags.  In
Fig.~\ref{fig1}c and \ref{fig1}d, $\tauN$ is as in LS outside $\scrD$,
but with $\tauN,\tauRB \rightarrow \infty$ inside $\scrD$ (i.e. the
Newtonian drag is not applied in $\scrD$ and the Rayleigh drag is not
applied anywhere in the domain).  Simulations in Fig.~\ref{fig1}a and
\ref{fig1}c are performed with BOB (T85L40 resolution; timestep size,
$\Delta t = 2\!\times\!  10^{-4}$; and, 8th-order hyperviscosity
coefficient, $\nu_8 = 1.5 \times 10^{-13}$), and simulations in
Fig.~\ref{fig1}b and \ref{fig1}d are performed with MITgcm using the
energy-conserving vorticity advection scheme (C64L40 resolution;
$\Delta t = 10^{-4}$; Shapiro filter order, $\mathfrak{n}\!  =\! 4$;
and, filter strength ratio, $\Delta t/\tau_{{\rm shap}}\! =\! 0.25$,
where $\tau_{{\rm shap}}$ is the Shapiro filter
coefficient).\footnote{Note that MITgcm can be sensitive to the
  numerical parameter values, as we discuss later; hence, simulations
  that match most closely with the BOB simulations are presented.}
Here `T85L40' denotes 40 vertical layers with 85 total and 85 sectoral
spherical harmonic modes in the Legendre expansion per layer, and
`C64L40' denotes 40 vertical layers with $6 \times 64 \times 64$ grid
points per layer \citep[see][for details of
the models]{Polietal14}.

Several features are readily apparent in Fig.~\ref{fig1}.  First, the
East and West BOB simulations in Fig.~\ref{fig1}a show no practical
difference between them, in agreement with LS using the MITgcm.
Second, not only do the East and West MITgcm simulations in
Fig.~\ref{fig1}b also show no practical difference between them, they
match very well with the corresponding BOB simulations in
Fig.~\ref{fig1}a.  The match is not exact, however, and we shall
return to this point shortly.  For now the salient point we wish to
make is that the MITgcm results here are nearly identical to those
presented in LS\footnote{cf. with Fig.~8 in LS, modulo plot aspect
  ratio and minor differences in the color range and palette}: hence,
there is no issue with the aforementioned minor difference in $T_e$ or
with the numerical parameter values chosen for the MITgcm simulations
in this study.  Third, as seen in Fig.~\ref{fig1}c and \ref{fig1}d,
when the strong drags are not imposed in $\scrD$, the
$\overline{u}^\ast$ distributions in the East and West simulations are
clearly distinct -- even when the data is heavily averaged, as in the
figures.  Grossly speaking, there are three jets in both, East and
West simulations, but the width, strength and vertical structure of
the jets are all noticeably different.  Observe as well that the East
and West simulations in Fig.~\ref{fig1}c (and Fig.~\ref{fig1}d)
separately do not match the corresponding East and West simulations in
Fig.~\ref{fig1}a (and Fig.~\ref{fig1}b) -- particularly below $\log(p)
\approx 3$ and away from the equatorial region.  According to the
figure, there are at least three plausible states, depending on the
setup and initial condition.

The first two features above confirm simultaneously the BOB
simulations in this study {\it and} the MITgcm simulations in LS.  The
very good agreement between the two models (cf. Fig.~\ref{fig1}a
and~\ref{fig1}b) is significant, as this is the first time extrasolar
planet atmospheric flow simulations from two different numerical
models are shown explicitly to produce nearly {\it quantitatively}
same results over the entire domain -- results which are numerically
converged over a good range of parameter values.  In general,
significant variations in the simulation results are observed across
different models, as well as within a single model \citep[see
e.g.][and the discussion below]{Polietal14,ThraCho11}.  It is
imperative to understand, however, that the good agreement here is due
to the highly constraining setup.  Nevertheless, as already pointed
out, the agreement is still not exact: the equatorial jets in
Fig.~\ref{fig1}a are faster and sharper than those in
Fig.~\ref{fig1}b, especially outside $\log(p)\!  \approx\!
[3.3,4.3]$.  LS report that in some cases modest variability is
exhibited but sensitivity is not exhibited, after the simulation
reaches the statistically steady state.  We have found that this is
the case with the setup as in Fig.~\ref{fig1}a and \ref{fig1}b.

The third feature confirms the behaviour reported in TC: the flow is
sensitive to the initial state.  In particular, it demonstrates
explicitly that the sensitivity observed in TC does not depend on
whether the day--night thermal gradient extends all the way down, or
only part way down, to the bottom of the domain.  LS speculate that
the sensitivity in TC may be caused by a global-scale baroclinic
instability induced by the gradient present at the bottom of the
domain.  However, TC have reported sensitivity in a variety of setup,
including one with reduced equilibrium temperature gradient at the
bottom.  Moreover, \citet{PoliCho12} have explicitly shown that, if a
flow is baroclinically unstable, the instability is not thwarted by
adding inactive layers (or drags) {\it below} the unstable region.
Hence, the placement of the bottom boundary is not the crucial factor
for sensitivity in these studies.  However, such factors (including
numerous other physical and numerical ones) can in fact affect
instabilities or wave dynamics that naturally arise in the course of
the flow evolution, steering it to different regions of the
configuration (solution) space -- and, this is so even if simulations
begin with identical initial conditions
\citep[e.g.][]{PoliCho12,ThraCho11}.  The point here is that the state
illustrated in Fig.~\ref{fig1}a (and \ref{fig1}b) is an extraordinary
one.

Interestingly, behaviour similar to that in Fig.~\ref{fig1}c and
\ref{fig1}d can also be seen in the correspondingly-similar
simulations of LS (see Fig.~15 therein, bottom row\footnote{Note, the
  Newtonian drag is employed in $\scrD$ here, unlike in
  Fig.~\ref{fig1}c and \ref{fig1}d; but, the inclusion of the
  Newtonian drag is not pertinent to the present discussion, as shown
  below.}).  But, LS interpret the behaviour as `artificial' because
the total angular momentum does not converge to the same value.  We
note here that there is no a~priori reason why this measure needs to
be the same among {\it different} model instantiations with different
initialisation and balance.\footnote{Here `balance' refers to the
  degree of vortical mode dominance (over gravity wave mode) of the
  flow field \citep[e.g.][]{Fordetal00}.}; in addition, even if the
measure is same, the spatial distribution of the flow (and thus
temperature) need not be same.\footnote{See \citet{Polietal14}; this
  can also be inferred from the East and West pairs of simulations in
  Fig.~\ref{fig2}b, presented below, in which the flow and temperature
  fields are markedly different despite starting with (and
  maintaining) same values of global angular momentum.}  Indeed, this
had been precisely TC's point and that of \citet{Choetal08}, as well
as the present paper: the (unknown) magnitude and distribution of
angular momentum and eddy kinetic energy in the actual atmosphere --
needed for model initialisation -- lead to divergent model evolutions,
preventing any kind of precise quantitative predictions at present.

At this point, we briefly discuss an important numerical issue: we
have found that MITgcm simulations suffer significant runaway angular
momentum and acute sensitivity to numerical parameters, when both
drags are not employed (Fig.~\ref{fig1}d).  An extensive study of
angular momentum conservation in several numerical models has been
performed and is described elsewhere (Politchouk \& Cho, in prep.).
The basic issue is illustrated in Fig.~\ref{fig2}, in which time
series of mass-weighted, global-average, axial angular momentum $M$
for simulations with Eastward-jet and Westward-jet initialisations at
different resolutions are presented\footnote{ $ M\ =\ (R_p/g)
  \int_{\mathfrak{V}} (\Omega R_p \cos\phi + u)\, \cos\phi\, {\rm
    d}V$, where $\Omega$ is the rotation rate of the planet, $R_p$ is
  the radius of the planet, $g$ is the gravity, $u$ is the zonal
  velocity, ${\rm d}V$ is the volume element and $\mathfrak{V}$ is the
  global domain.}; each series is normalised by its initial value,
$4.34\!\times\! 10^{34}$\,kg\,m$^2$\,s$^{-1}$ for the East simulations and
$2.76\!\times\! 10^{34}$\,kg\,m$^2$\,s$^{-1}$ for the West simulations.
Fig.~\ref{fig2}a shows the runaway and sensitivity (with resolution)
in MITgcm simulations.  The runaway and sensitivity reduce at high
resolution (e.g. at C64), but do not vanish.  In contrast, runaway and
sensitivity are not observed in BOB simulations, as seen in
Fig.~\ref{fig2}b.  The runaway is detectable in Fig.~\ref{fig1}:
notice the small north--south hemispherical symmetry breaking in
Fig.~\ref{fig1}d, but not in Fig.~\ref{fig1}c.  Since external
symmetry breaking has not been introduced in these simulations, the
asymmetry is a manifestation of the runaway -- i.e. a numerical error.
Hence, total angular momentum is not a useful measure for MITgcm in
this case; and, BOB and CAM results (which are in good agreement with
each other) cannot be robustly reproduced or tested with MITgcm when
strong drags are not applied in $\scrD$.

\begin{figure}
  \centerline{\includegraphics[scale=0.47]{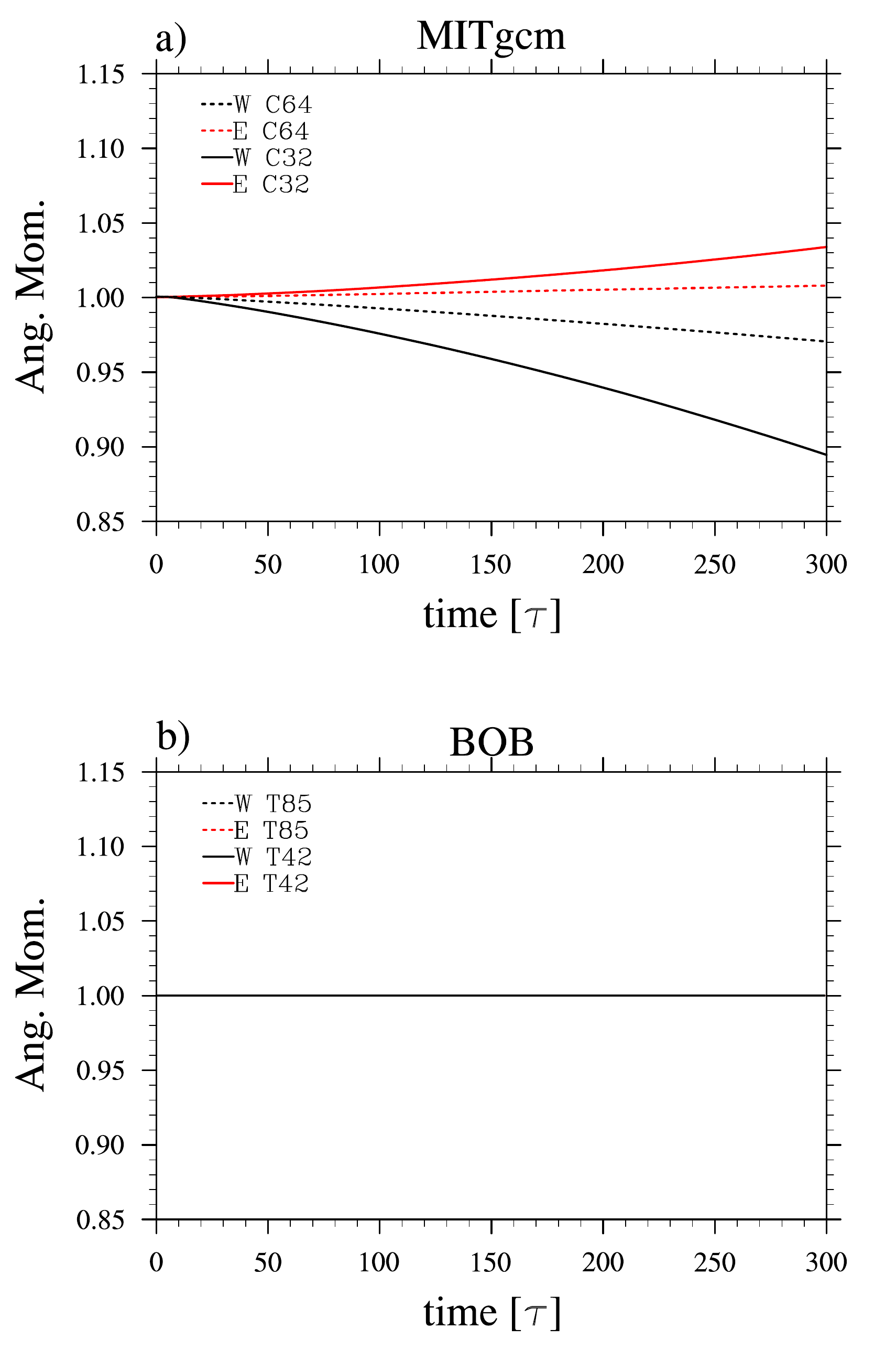}}
  \caption{Mass-weighted, global-average, axial angular momentum
    [kg\,m$^2$\,s$^{-1}$] time series for simulations with
    East(E)\,/\,West(W) initialisations at different resolutions.
    Each series is normalised to its initial value: $4.34\!\times\!
    10^{34}$\,kg\,m$^2$\,s$^{-1}$ (E) and $2.76\!\times\!
    10^{34}$\,kg\,m$^2$\,s$^{-1}$ (W) simulations.  The physical setup
    is as in Fig.~\ref{fig1}d (MITgcm) and Fig.~\ref{fig1}c (BOB).
    MITgcm exhibits monotonic runaway behaviour, which reduces with
    increased resolution.  In contrast, the BOB model conserves
    angular momentum exactly, independent of the resolution.}
  \label{fig2}
\end{figure}

\begin{figure*}
  \centerline{\includegraphics[scale=0.7]{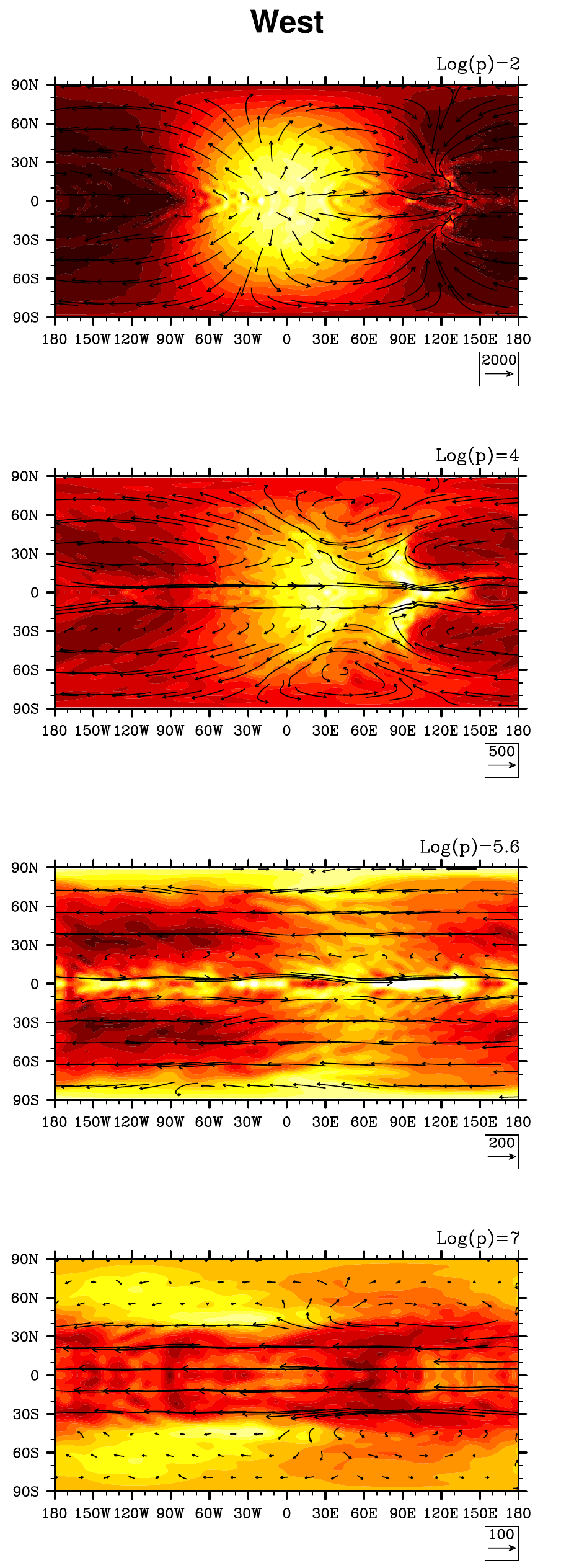} \qquad \quad
    \qquad
     \includegraphics[scale=0.7]{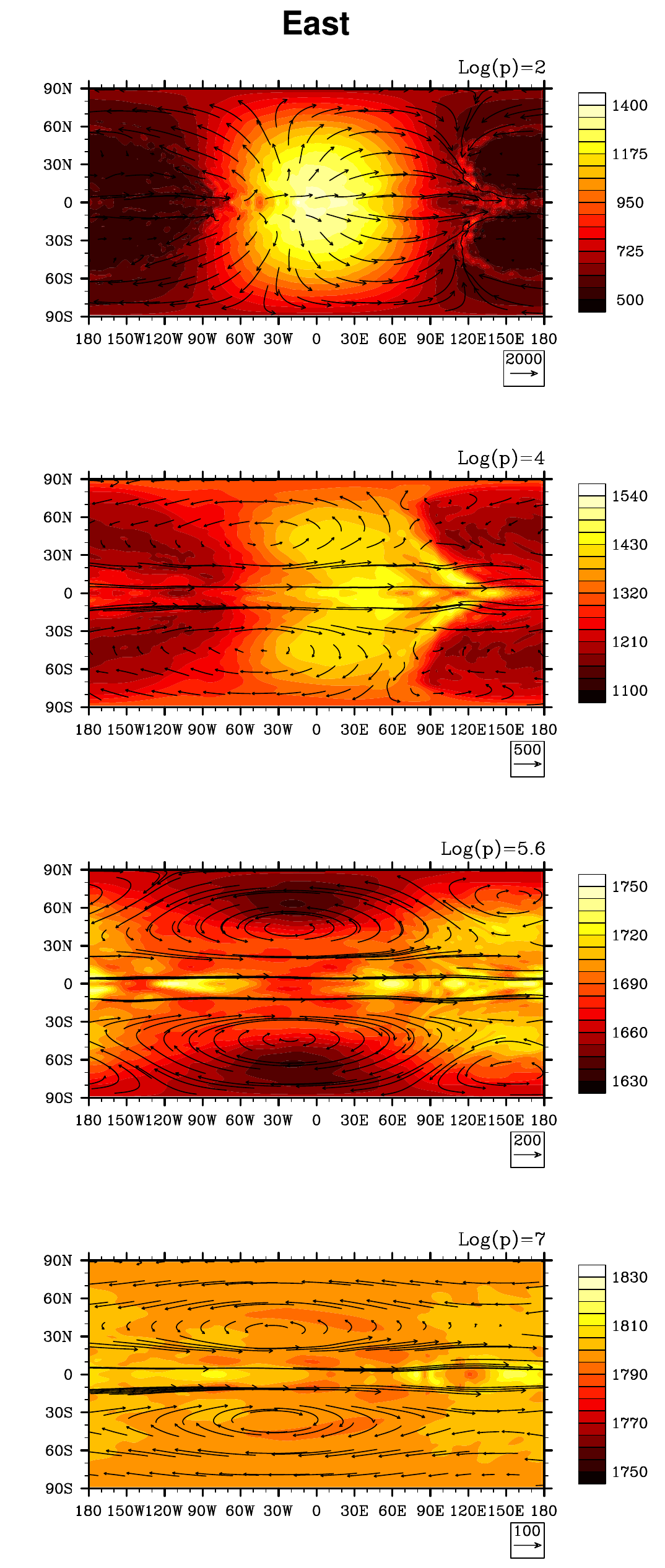}}
   \caption{Instantaneous latitude--longitude maps of the flow ${\bf
       v}$~(m~s$^{-1}$) and temperature T~(K) fields at four pressure
     levels, $\log(p)\! =\!  \{2.0, 4.0, 5.6, 7.0\}$, at $t\! =\!
     300$ in cylindrical-equidistant projection from the (West/East)
     simulations in Fig.~\ref{fig1}c; ${\bf v} = (u,v)$, where $u$ and
     $v$ are the zonal and meridional velocities, respectively.  The
     reference flow vectors are shown at the bottom right in each
     panel and temperature ranges for each row are shown at the right.
     In both West and East simulations, the flow and temperature
     distributions are complex with multiple, irregular hot/cold
     regions -- often situated away from the equator and
     substellar/antistellar (0$^\circ$/180$^\circ$) longitude.  The
     spatial distribution of the flow and temperature fields are
     significantly different, except near the top of the modelled
     domain (see Fig.~\ref{fig4}).  Symmetry has not been broken
     initially, unlike in most simulations in TC.  The vertical
     temperature gradient fields in the two simulations are also
     different, and would lead to different emergent heat flux
     distributions.}
  \label{fig3}
\end{figure*}

Given the zonal and temporal averaging in Fig.~\ref{fig1}, it may be
difficult to fully appreciate just how different the fields can be in
simulations initialised with different jets when the strong drags are
not applied in $\scrD$.  Unaveraged fields contain more information
and are more useful for comparing with observations.  Fig.~\ref{fig3}
shows the instantaneous flow and temperature fields from the East and
West simulations of Fig.~\ref{fig1}c at the levels, $\log(p)\!  =\!
\{2.0,4.0,5.6,7.0\}$, at $t\!  =\! 300$.  Both the flow and
temperature distributions are markedly different in the two columns
(East and West simulations), except near the top of the domain
(cf.~$\log(p)\! =\! 2$ frames).  The behaviour near the top is
expected, given the extremely strong Newtonian drag applied there
($\tauN\approx 0.03$).  But, even there significant differences are
present in both fields, as will be shown more clearly below.

Significantly, there is much emphasis in the current literature on the
temperature distribution at high altitude (e.g. $p\!  =\!  30$~mbar $
\Rightarrow \log(p)\!\approx\!  3.5$), where both the temperature and
the flow are highly restrained by the very short $\tauN$ ($\approx\!
0.1$).  However, attention should also be given to the lower regons --
for a more realistic, complete picture.\footnote{This also applies to
  the mid- and high-latitude regions.}  First, note that such
timescales are comparable to the Brunt--V\"{a}is\"{a}l\"{a} period,
the timescale of `deep' internal gravity waves.  The distortion or
omission of various waves, including the gravity waves, is a source of
significant inaccuracy in current atmospheric flow simulations: for
example, wave interactions with the background flow (which modify the
temperature distribution) are poorly captured -- if at all \citep[see
e.g.][and also this study below]{Fordetal00,Choetal03,WatCho10}.
These interactions can be long-range and be effected by waves
originally of small amplitude or scale.  Second, infrared flux itself
can originate from any of the levels modelled, consistent with the use
of Newtonian relaxation approximation.  The approximation is meant to
represent crudely the effect of radiation emanating from the
atmospheric region where the approximation is being invoked
\citep[e.g.][]{Salby96, Andetal87,Choetal08}.  Given the above,
matches with observations in current simulations may merely be
fortuitous: a better understanding of both the physics and numerics is
needed.

\begin{figure*}
  \centerline{\includegraphics[scale=0.66]{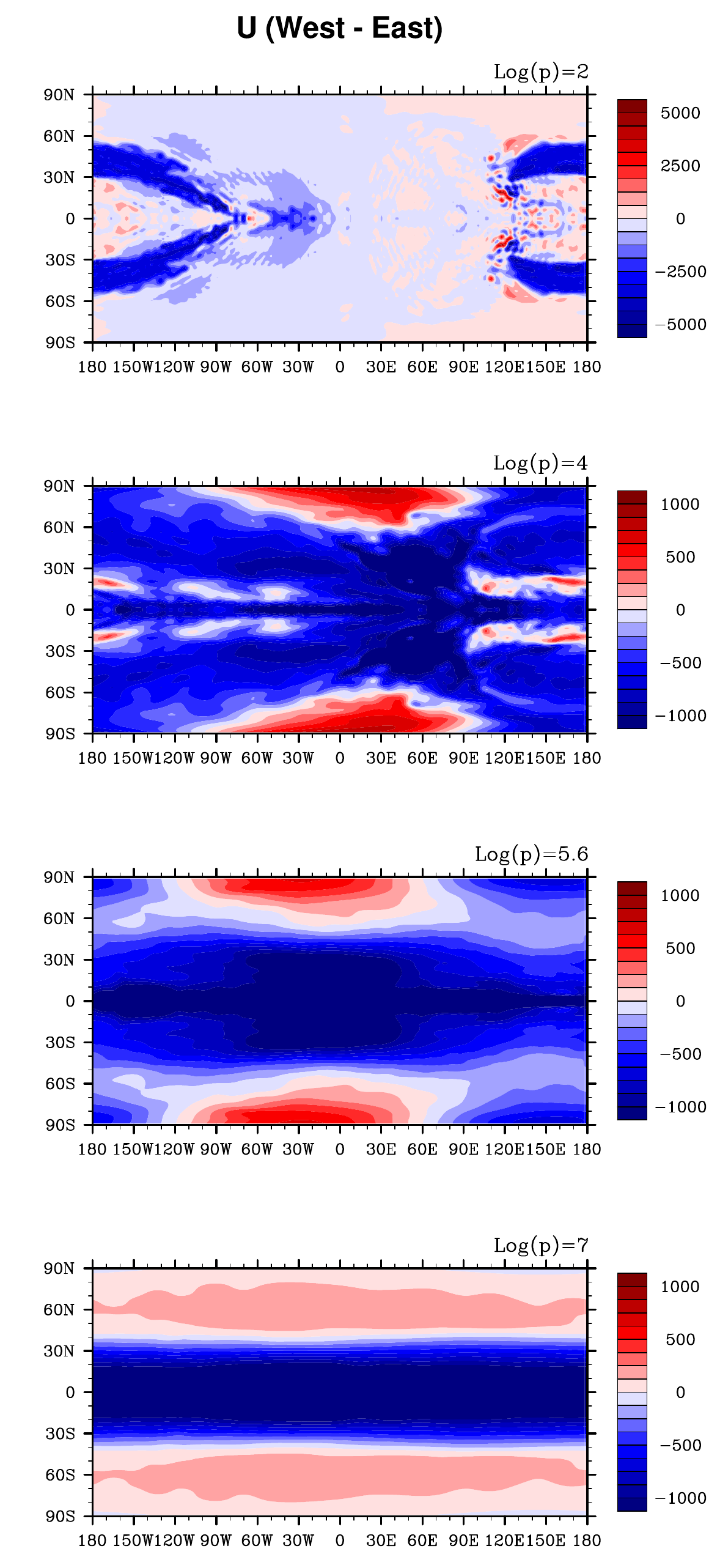} \qquad
    \includegraphics[scale=0.66]{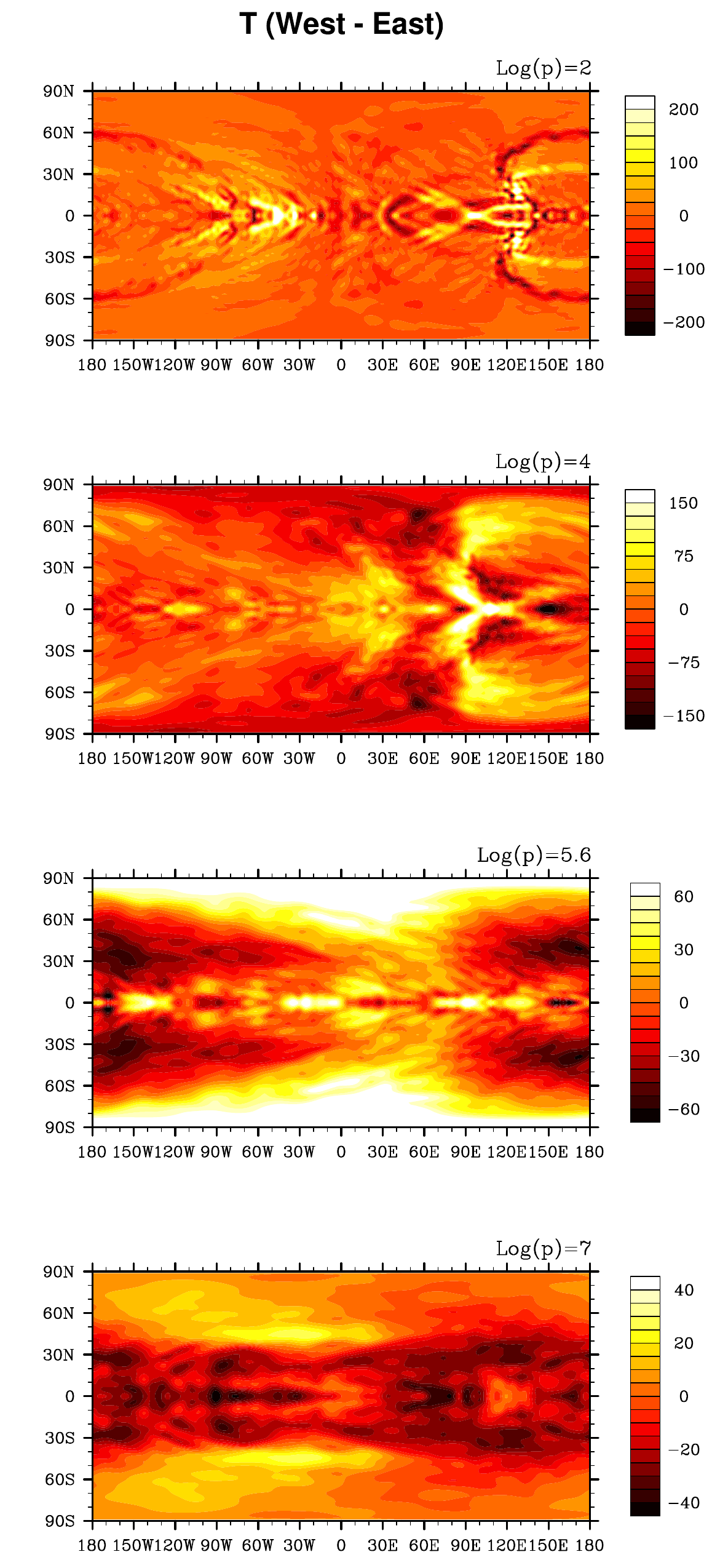}}
  \caption{{\it West} $-$ {\it East} difference maps of the zonal wind
    $u$~(m~s$^{-1}$) and temperature $T$~(K) fields at the $p$-levels
    in Fig.~\ref{fig3}.  The wind differences are at the left column,
    and the temperature differences are at the right column.  The plot
    ranges are shown at the right of each panel.  Locally, the
    absolute flow and temperature differences can be large -- as much
    as $\sim$4000~m~s$^{-1}$ and $\sim$200~K at $\log(p) = 2$,
    respectively.}
  \label{fig4}
\end{figure*}

Note also that the fields in Fig.~\ref{fig3} are very similar to those
in TC, despite the bottom boundary being located at a much greater
depth in the simulations of the figure.  As in TC, the hottest and
coldest regions are not simply connected \citep[e.g.][]{Munk00} and
often located away from the equator.  This is in contrast to `maps'
constructed from observations which assume a latitudinal distribution
monotonically decaying away from the equator.  Moreover, the hottest
(coldest) region can often be situated at the night (day) side, also
in contrast to what has been reported in some observations and
simulations \citep[e.g.][LS, and references therein]{Knutsonetal07}.
In general, the locations of the temperature extremum regions vary in
space and time, within a single simulation and across different
simulations \cite[e.g.][]{Choetal03,Choetal08,ThraCho10,Polietal14},
again depending on the initial flow and equilibrium temperature states
specified.

For a clearer picture of the magnitude and spatial distribution of the
{\it differences}, Fig.~\ref{fig4} presents point-wise subtractions of
the West and East frames at the $p$-levels shown in Fig.~\ref{fig3}.
In Fig.~\ref{fig4}, the left column shows the instantaneous zonal wind
field difference, $u_{\mbox{\tiny W}}(\lambda,\phi) - u_{\mbox{\tiny
    E}}(\lambda,\phi)$, and the right column shows the instantaneous
temperature field difference, $T_{\mbox{\tiny W}}(\lambda,\phi) -
T_{\mbox{\tiny E}}(\lambda,\phi)$.  Note the large, absolute maximum
flow and temperature differences (more than $\sim$4000~m~s$^{-1}$ and
$\sim$200~K, respectively) locally at $\log(p)\! =\! 2$.  At greater
depths the $u$--differences are reduced, but are still quite large
($\sim$1000~m~s$^{-1}$) and on global scales.  The $u$--difference in
the left bottom panel reflects the opposite sign of the zonal jets in
the $\scrD$ region in the two simulations, particularly in the
equatorial region (cf. East and West simulations in Fig.~\ref{fig1}c,
for example).  The $T$--differences also generally decrease with
depth.  Notably, in the $\scrD$ region, the temperature difference is
a very small fraction of the ambient temperature ($\!\lsim 4$\%) in
both of the simulations, indicating that the Newtonian drag as
specified in simulations of Fig.~\ref{fig1}a and \ref{fig1}b is not
really effective or needed.  This is not surprising, given the large
inertia of $\scrD$.

\begin{figure}
  \centerline{\includegraphics[scale=.55]{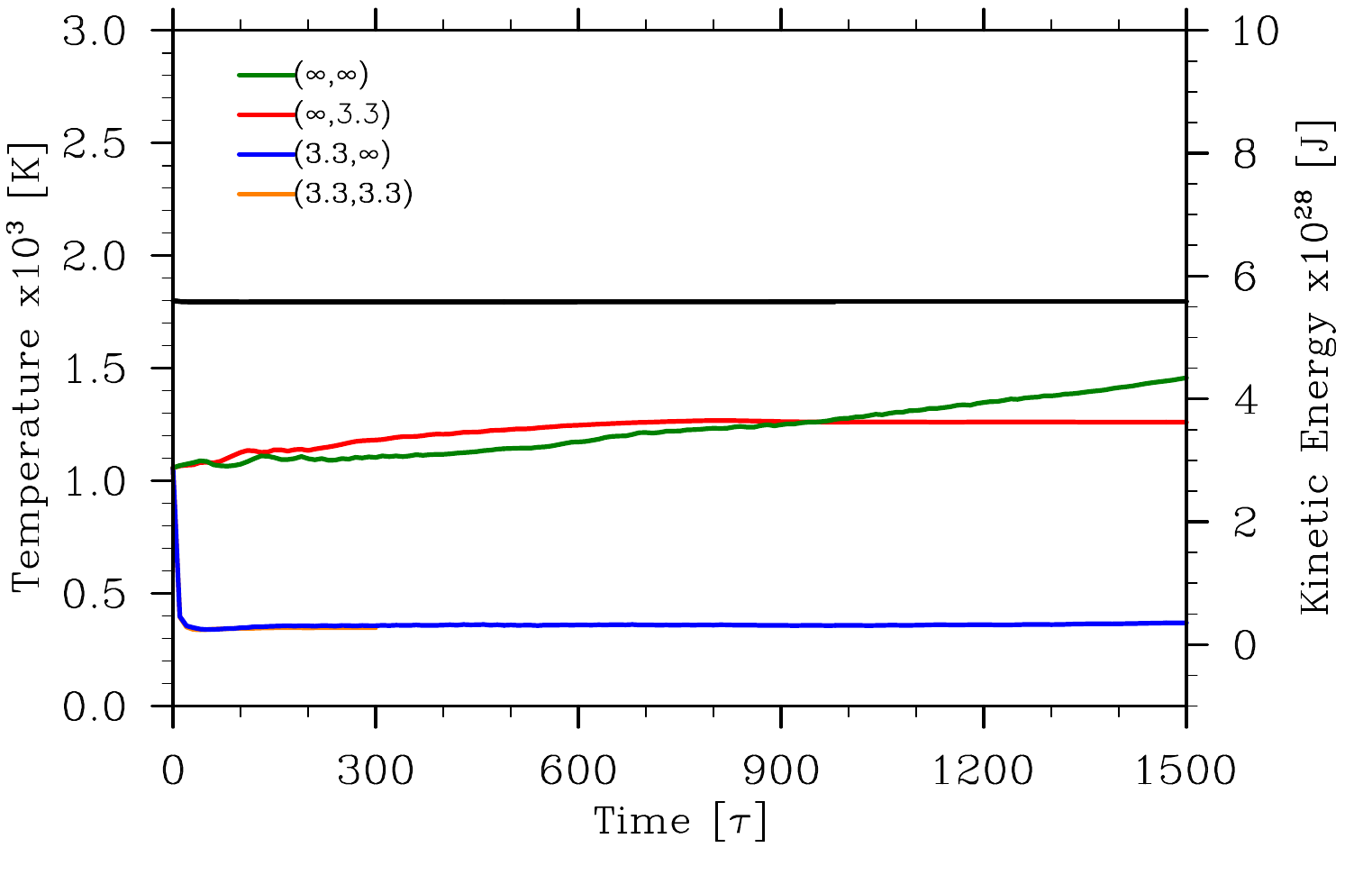}}
  \caption{Mass-weighted, global-average temperature $\langle T
    \rangle$ (black) and kinetic energy $\langle \mathscr{K} \rangle$
    (green, red, blue and orange) time series from four simulations
    set up identically, except for different $(\tauRB,\tauN)$ in
    $\scrD$.  When finite, $\tauR(p)$ decreases linearly with $p$ in
    $\log(p) = [5,7.3] = [5,6]\!\cup\!\scrD$ and $\tauN$
    is constant in $\scrD$; outside their respective regions,
    $\tauN(p)$ and $\tauR(p)$ are as in LS in all cases.  All are
    T85L40 BOB simulations, initialised with the $+1000$~m~s$^{-1}$
    jet and ${\cal T}_0\! =\! 1800$~K.  The black curve is common to
    all four simulations.  The blue and orange curves are very close
    to each other.  The green and red curves are different for a
    different initialisation while the blue and orange curves are
    not.}
  \label{fig5}
\end{figure}

In their study, LS motivate the use of a strong Rayleigh drag in as an
`accelerator' to help reach the `equilibrium' state more quickly.
However, such a drag can steer the flow to artificial states and is in
actuality unnecessary -- particularly if temperature is used to
characterise the equilibration, as in many climate studies
\citep[e.g.][]{BranOtto09}, or if strong Newtonian drag is already
applied in $\scrD$ (which is also dubious, as noted above).  Both can
be clearly seen in Fig.~\ref{fig5}, where time series normalised by
the initial value from four long-duration simulations (roughly
`permutations of $\{(\tauRB,\tauN)\}$') are presented; all the
simulations here employ strong Newtonian drag in the upper region.  In
the figure, the mass-weighted, global-average temperature $\langle T
\rangle$ is steady in all the simulations presented, regardless of the
drags employed in $\scrD$ (black curve).  With only the Newtonian drag
applied in $\scrD$, the mass-weighted, global-average kinetic energy
$\langle \mathscr{K} \rangle$ reaches steady state (at $t\!\approx\!
700$, red curve).  With both drags not applied in $\scrD$, slow
increase in $\langle \mathscr{K} \rangle$ is observed (green curve),
but the rate of increase depends on time (and on the model setup).  We
stress that, in principle, such an evolution is physically valid -- as
long as $\langle \mathscr{K} \rangle$ does not `blow up'.  Observe
that the Rayleigh drag is responsible for suppressing the sensitivity
(cf.~orange, blue and red curves): for this, Newtonian drag by itself
is ineffectual.

LS also suggest, following e.g. the study by \citet{Pernaetal10} for
higher altitude, that a Rayleigh drag in $\scrD$ might serve as a
crude representation of `magnetic drag effects' stemming from thermal
ionisation. There are two major concerns with this.  First, thermal
ionisation is insignificant in the $\scrD$ region: temperature is too
low and density is too high.  This is so even taking into account the
low ionisation potential of alkali metals (e.g. K, Na, Ca), as these
are trace species \citep[see e.g.][]{Lewis04}.  Using solar
abundances, $n_{{\rm H}^+}/n_{\rm n}\lesssim 2\times10^{-16}$ and
$n_{{\rm K}^+}/n_{{\rm H}^+} \approx 3\!\times\!  10^6$, where $n_x$
is the $x$-specie number density and `n' subscript refers to {\it
  neutral}.  Hence, the bulk ionisation level (electron volume mixing
ratio), $\chi_{\rm e} \equiv n_{\rm e}/n_{\rm n}\approx n_{{\rm
    K}^+}/n_{\rm n}$, is at least $10^2$ times lower than that
required for the fluid medium to be influenced electromagnetically:
$\chi_{\rm e}\!  \sim\!  10^{-7}$ is required for ionisation drag
effects to become significant \citep[e.g.][]{SchuNagy00,Kosketal14}.
Second, even if the ion-induced drag were significant via a
non-thermal mechanism, it cannot be represented as an {\it isotropic
  drag to rest} on the momentum\footnote{This is also true of gravity
  wave induced drag.}: ion velocities and the intrinsic field
orientation need to be modeled self-consistently for accurate
representation \citep[e.g.][]{Kosketal10}.

\begin{figure*}
  \includegraphics[height=4.8cm,width=17.8cm]{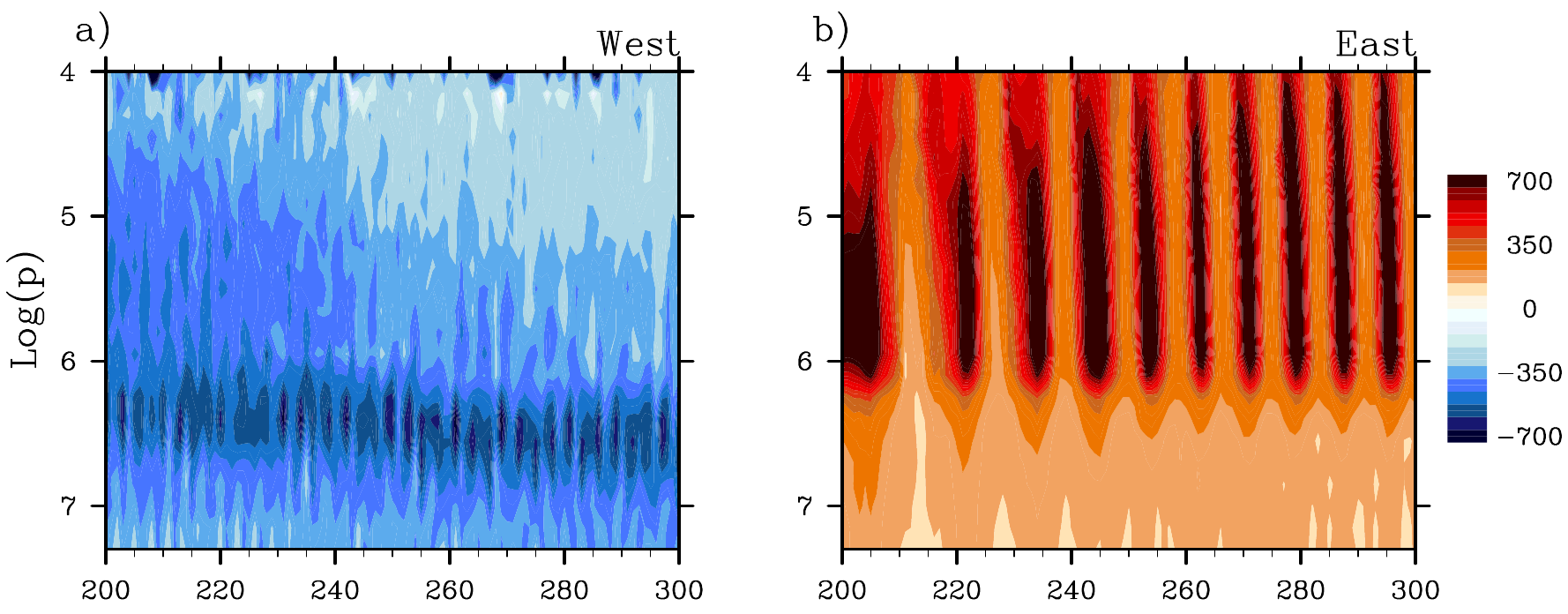}
  \vspace*{.5cm}
  \includegraphics[height=5.3cm,width=17.8cm]{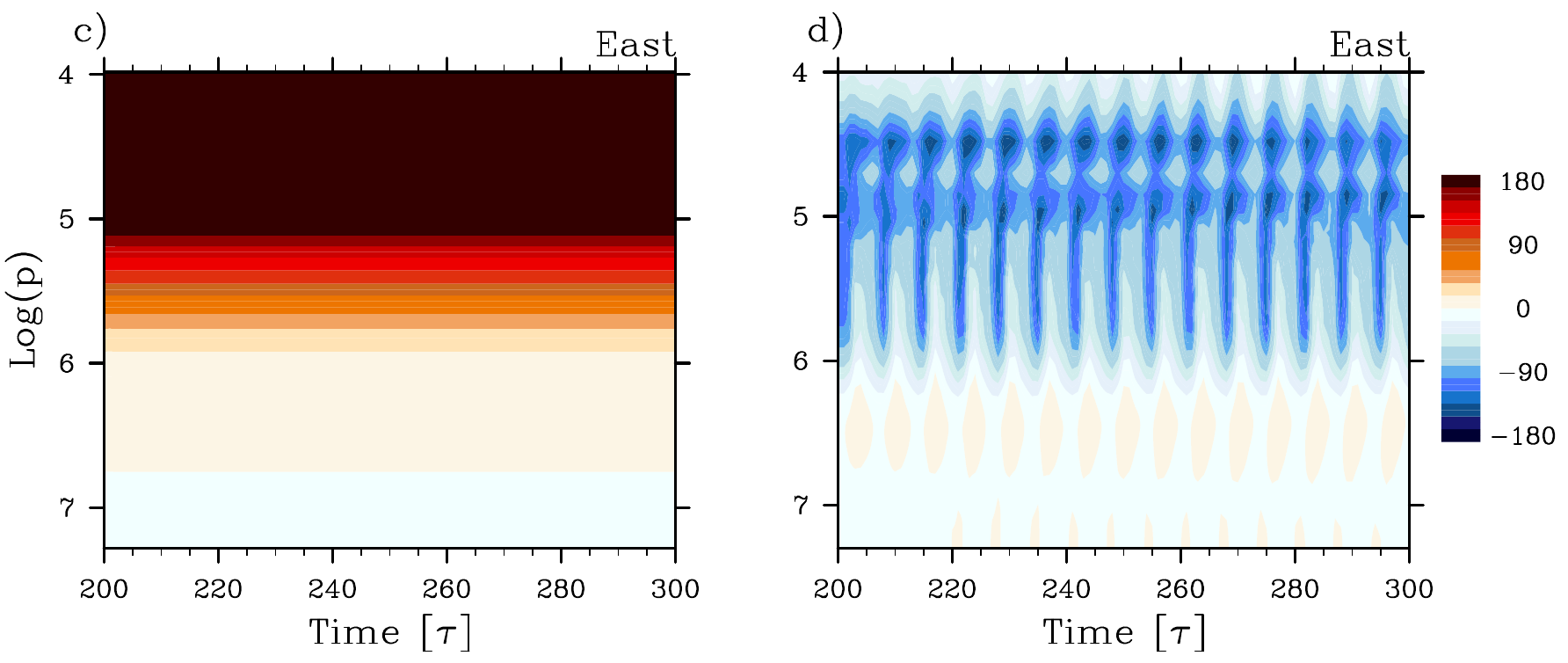}
  \caption{$t$--$p$ Hovm\"oller plots of instantaneous zonal velocity
    $u(t,p)\vert_{(\lambda,\phi) = (0,30)}$ for $t = [200,300]$ and
    $\log(p) = [4.0,7.3]$.  The colour bars indicate the velocity for
    each row.  All simulations are performed with BOB.  Initial jet
    amplitudes and directions (indicated above each plot) are as in
    Fig.~\ref{fig1}.  In a) and~b), the resolution is T85L40 and
    $\tauRB,\tauN \rightarrow \infty$ in $\scrD$.  In c) and d), the
    resolution is T21L1000 and $\tauRB,\tauN \rightarrow 3.3$ in
    $\scrD$ (as in LS) with the vertical levels equally-spaced in
    $\log(p)$ (c) and $p$ (d).  Vertically-propagating waves of
    planetary scale power variability and behave differently,
    depending on the initial flow, when strong drags are not applied
    in $\scrD$ [cf. a) and b)].  However, even with the strong drags
    employed, propagating waves are present if the vertical resolution
    is increased in $\scrD$ [cf. c) and d)].}
  \label{fig6}
\end{figure*}

Let us now consider a more detailed look at the behaviour in time.
Fig.~\ref{fig6} presents $t$--$p$ Hovm\"oller plots of the zonal
velocity, $u(\lambda\! =\! 0^\circ\!, \phi\! =\!  30^\circ\!, p,\,
t)$, for $t = [200,300]$ and $\log(p) = [4.0,7.3]$.  Fig.~\ref{fig6}a
and \ref{fig6}b are from the simulations in Fig.~\ref{fig1}c, which is
at T85L40 resolution.  Their general behaviours have been verified
with up to T341L40 resolution using the BOB model.  Fig.~\ref{fig6}c
presents a simulation with a setup identical to that of the simulation
in Fig.~\ref{fig1}a (East), except the resolution here is T21L1000
($\Delta t = 8\!\times\!  10^{-4}$ and $\nu_8 = 3 \times 10^{-10}$).
The results are nearly identical at T42L500 and T85L40 resolutions.
The simulation at T21L1000 resolution is presented here for equatable
comparison with the simulation shown in Fig.~\ref{fig6}d, in which the
setup is identical to the simulation shown in Fig.~\ref{fig6}c, except
the vertical layers are now equally spaced in $p$.  This spacing has
the effect of representing the lower (higher) region with greater
(fewer) number of layers: the resulting resolution in
Fig.~\ref{fig6}d, for example, is 57 times greater at the bottom of
$\scrD$ than that in Fig.~\ref{fig6}c and in LS.  In the simulation
presented in Fig.~\ref{fig6}d, the spacing is more numerically {\it
  consistent}\footnote{Numerical consistency refers to discretisation
  error tending to zero as the resolution is increased, under
  pointwise convergence at each grid point \citep[e.g.][]{Stri04}.},
as the equations solved in all the simulations are in $p$ rather than
$\log(p)$ coordinate.

Fig.~\ref{fig6}a and \ref{fig6}b show vertically-propagating Rossby
(planetary) waves, generally prominent in the model atmospheres when
strong drags are not applied in $\scrD$.  In the two panels, the waves
are propagating in opposite directions (upward and downward in
Fig.~\ref{fig6}a and \ref{fig6}b, respectively), as evident from the
tilt of the phase lines.  The magnitude of the peak amplitude is
$\sim$700~m~s$^{-1}$ in both, given the latitude location of the
constructed plots, but the sign and period are different -- as is the
growth with height over time.  As already discussed, such waves can
induce significant modification of the background flow via saturation
and encounter with critical layers \citep[e.g.][]{Andetal87,Holt04}.
The action of the waves will be discussed more in detail elsewhere.
Here we wish to highlight the suppression of essentially {\it all}
temporal activity -- including these waves -- when the strong drags
are applied in $\scrD$, in addition to a strong Newtonian drag already
applied throughout nearly all of the domain outside $\scrD$
(cf. Fig.~\ref{fig6}c with Fig.~\ref{fig6}b).  While not
uninteresting, such a `dead' atmosphere shown in Fig.~\ref{fig6}c is
not dynamic, and a flow simulation of it is not very informative.

Interestingly, the suppressed behaviour is not robust and appears to
be a numerical artifact.  This is demonstrated in Fig.~\ref{fig6}d.
The panel shows the behaviour of the simulation in Fig.~\ref{fig6}c
when the vertical resolution of the dynamically active region
(\,$\log(p) \approx [5,6]$\,) is increased, principally by employing a
different spacing.  As can be seen in Fig.~\ref{fig6}d, the
variability returns -- powered by waves that strongly shear, emerging
from the $\scrD$ region, and induce secondary and even tertiary waves
upon interacting strongly with the northern flank of the equatorial
jet at $\log(p) \approx [5.0, 4.5]$ (see Fig.~\ref{fig1}a).  The
amplitude of the waves here is much smaller than those in
Fig.~\ref{fig6}a and \ref{fig6}b, consistent with the much stronger
drag in this case.  Propagating gravity waves that are similarly
generated can enhance the observed variability as well
\citep[e.g.][]{SciFord00, WatCho10}, but these waves are poorly
captured in this simulation at the employed resolution.

\section{Coda}\label{coda}

We close with some thoughts and observations arising from this study.
We have carefully cross-checked the results with codes used in TC and
LS, as well as a third code (which has been extensively tested under
conditions appropriate for hot-Jupiters).  As shown by
\citet{Polietal14} and the present study, one can arrive at erroneous
conclusions if simulations from different codes are not at least
qualitatively reproduced {\it with the same setup}.  Even then,
erroneous conclusions can still be drawn, as codes which perform well
ostensibly in one region of the physical parameter space do not
perform well in another (or, more precisely, extended) region.  For
example, when pushed to a highly ageostrophic\footnote{e.g. high
  Rossby and Froude numbers -- measures of rotation rate and
  stratification, respectively \citep[see e.g.][]{Holt04}} region (as
would occur in a typical hot-Jupiter atmosphere simulation), numerical
accuracy of the code can become seriously degraded by small-scale
oscillations generated in that region of the parameter space
\citep[e.g.][]{ThraCho11}.

By expanding the study by \citet{Polietal14} into the {\it initial
  condition space}, the results here constitute an extension of that
study: in general, simulations are sensitive to initial condition --
in addition to many other factors, physical and numerical.  That the
physical system under study exhibits multiple equilibrium states
(hence, also non-ergodicity) is not, it seems to us, a particularly
startling assertion given its highly nonlinear and poorly constrained
forced-dissipative nature.  But, it has been challenged forcefully by
LS.

Ultimately, it appears the question of sensitivity and variability
rests on whether one believes that it is truly `realistic' to apply
Rayleigh drag as well as Newtonian drag on very short timescales,
particularly in the $\scrD$ region.  Even assuming that the drags are
acceptable representations of the forcing and dissipation in
hot-Jupiter atmospheres, one must ask: how realistic are $\tauN$,
$\tauR$ and~$T_e$ currently used?\, The responses of TC and LS clearly
differ on this question.  A telling observation of this study is the
extraordinary length one must go to suppress sensitivity and
variability inherent in the system: {\it all eddies and waves must be
  eliminated for all time}.  One may wonder whether such a state is
really realistic, given that the atmosphere would likely contain some
vorticity and turbulence (both inhomogeneous and intermittent) and
that there are altitude regions (e.g. just above $\scrD$)
characterised by neither short Newtonian nor short Rayleigh drag
timescales.  In our view, the general question of realistic forcing
and initialisation -- and their effects -- is still unsettled, and the
question deserves much more attention and scrutiny than has generally
received thus far.

Most importantly, the investigation should proceed with the following
observations squarely in the fore.  We have found that MITgcm in the
cubed-sphere grid configuration conserves angular momentum poorly and
is acutely sensitive to numerical parameter values in simulations
initialised with high speed jets, particularly without the strong
Rayleigh drag applied in $\scrD$ (e.g. Fig.~\ref{fig2}): hence, it
cannot be used for assessing sensitivity in this case, as it is not
possible to establish a reliable baseline for quantification.  We have
also found that, {\it all} codes used in this study (including BOB)
lead to a zonally-symmetric, superrotating, supersonic jet at the
equator, at $\log(p) \approx 3$.  Given that the hydrostatic primitive
equations with free-slip boundary conditions (as well as with the
viscosity representation for small Mach number flows) are solved in
all the numerical models in this study, such flow is physically not
valid [see e.g. discussion in \citet{Holt04}].  Therefore, claims of
realism in these setups is moot: such jets may occur on real
hot-Jupiters, but not in these simulations.  Either a different setup
must be used for the equations and boundary conditions solved {\it or}
a different set of equations and boundary condition must be solved for
the setup used.  Relatedly, while all of the simulations presented
{\it here} exhibit supersonic zonally-symmetric equatorial flow at
some height, not all simulation do: in general, that depends on the
physical setup, numerical scheme and initial condition employed
\citep[e.g.][]{ThraCho10,Polietal14}.

\section*{Acknowledgments}

The authors acknowledge helpful discussions with Craig Agnor, Tommi
Koskinen and Stephen Thomson -- and particularly for their careful
reading of the manuscript.  The authors also acknowledge the
hospitality of the Kavli Institute for Theoretical Physics, Santa
Barbara, where some of this work was completed.  I.P. is supported by
UK's Science and Technology Facilities Council research studentship.
H.Th.Th. is supported by the NASA Post-doctoral Program at the Jet
Propulsion Laboratory, administered by Oak Ridge Associated
Universities through a contract with NASA.

\end{document}

%% file: chocros.tex
\usepackage{amssymb}
\usepackage{upgreek}
\usepackage{bbm}
\usepackage{dsfont}
\usepackage{sansmath}
\usepackage{mathrsfs}
\usepackage{yfonts}
\usepackage{euscript}

\def\grad-s{\mbox{\boldmath{$\nabla$}}_{\!\! s}\,}

\def\tauN{\tau_{\mbox{\tiny{\sc N}}}}
\def\tauR{\tau_{\mbox{\tiny{\sc R}}}}
\def\tauRB{\tau_{\mbox{\tiny{\sc R20}}}}

\def\scrD{\mathscr{D}}

\def\gsim{\;\rlap{\lower 2.5pt\hbox{$\sim$}}\raise 1.5pt\hbox{$>$}\;}
\def\lsim{\;\rlap{\lower 2.5pt\hbox{$\sim$}}\raise 1.5pt\hbox{$<$}\;}

\usepackage{color}
\definecolor{com_red}{rgb}{.8,.2,0.1}
\definecolor{ins_green}{rgb}{.1,.5,0.1}